\documentstyle [12pt] {article}

\newcommand{\beq}{\begin{equation}}
\newcommand{\eeq}{\end{equation}}
\newcommand{\beqn}{\begin{eqnarray}}
\newcommand{\eeqn}{\end{eqnarray}}
\newcommand{\eq}[1]{(\ref{#1})}
\newcommand{\bi}[1]{\bibitem{#1}}
\newcommand{\fr}[2]{\frac{#1}{#2}}
\newcommand{\dst}{&\displaystyle}

\newcommand{\al}{Z\alpha}
\newcommand{\alv}{\mbox{\boldmath $\alpha$ \unboldmath}}

\newcommand{\e}{\mbox{${\bf e}$}}
\newcommand{\eps}{\varepsilon}

\newcommand{\gv}{\mbox{\boldmath $\gamma$ \unboldmath}}
\newcommand{\k}{\mbox{${\bf k}$}}

\newcommand{\p}{\mbox{${\bf p}$}}
\newcommand{\pv}{{\mathbf p}}
\newcommand{\lv}{{\mathbf L}}
\newcommand{\Q}{\kappa}
\newcommand{\r}{{\mathbf r}}
\newcommand{\R}{{\mathbf R}}
\newcommand{\ro}{\mbox{\boldmath ${\rho}$\unboldmath}}

\newcommand{\vt}{\mbox{\boldmath ${\theta}$\unboldmath}}

\newcommand{\vx}{{\bf x}}

\newcommand{\ls}{\leq}
\newcommand{\gs}{\geq}

\newcommand{\q}{{\bf q}}
\newcommand{\vq}{\mbox{${\bf Q}$}}
\newcommand{\vd}{\mbox{${\bf \Delta}$}}

\newcommand{\grad}{\mbox{\boldmath ${\nabla}$\unboldmath}}

\begin {document}
\begin{titlepage}
\begin{center}
{\Large \bf Budker Institute of Nuclear Physics}
\end{center}

\vspace{1cm}

\begin{flushright}
{\bf Budker INP 99-63\\
July 27, 1999 }
\end{flushright}

\vspace{1.0cm}

\begin{center}{\Large \bf Quasiclassical Green function in an
external field and small-angle scattering}
\end{center}
\vspace{1.0cm}

\begin{center}
{{\bf R.N. Lee, A.I. Milstein, V.M. Strakhovenko}\\
G.I. Budker Institute of Nuclear Physics, \\
 630090 Novosibirsk, Russia}
\end{center}
\vspace{3cm}

\begin {abstract}
The quasiclassical Green functions of the Dirac and Klein-Gordon
equations in the external electric field are obtained with the first
correction taken into account. The relevant potential is assumed to be
localized, while its spherical symmetry is not required. Using these
Green functions, the corresponding wave functions are found in the
approximation similar to the Furry-Sommerfeld-Maue approximation. It
is shown that the quasiclassical Green function does not coincide
with the Green function obtained in the eikonal approximation and has
a wider region of applicability. It is illustrated by the calculation
of the small-angle scattering amplitude for a charged particle and the
forward photon scattering amplitude. For charged particles, the first
correction to the scattering amplitude in the non-spherically
symmetric potential is found. This correction is proportional to the
scattering angle. The real part of the amplitude of forward photon
scattering in a screened Coulomb potential is obtained.
\end {abstract}

\end{titlepage}

\section {Introduction}
As known, the use of the wave functions and Green functions of the
Dirac equation in an external field is a convenient tool for the
calculation of QED amplitudes in the field. At high energy and small
scattering angles, the characteristic angular momenta of
involved particles are large, and the quasiclassical approximation is
applicable.  In this case the use of quasiclassical Green functions
greatly simplifies the calculations.

The quasiclassical Green function $G(\r_2,\r_1|\eps)$ of the Dirac
equation in a Coulomb potential was first derived in \cite{MS1, MS2}
starting from the exact Green function of the Dirac equation
\cite{MS0}. The more convenient representation of the quasiclassical
Coulomb Green function for the case of almost collinear vectors $
\r_1 $ and $ \r_2 $ was obtained in \cite{LMS1, LMS2}. In the same
geometry the quasiclassical Green function in arbitrary spherically
symmetric decreasing potential was found in \cite{LM1, LM2}.

In section 2 we derive the quasiclassical Green functions of the
Dirac and Klein-Gordon equations for the case of a localized
potential not assumed to be spherically-symmetric. We
use the term 'localized potential' for the decreasing potential
having a maximum at some point. The Green functions are obtained with
the first correction taken into account.  With their help the
quasiclassical wave functions of the Dirac and Klein-Gordon equations
and corrections to them are found (section 3).  These wave functions
generalize the results obtained in the Furry-Sommerfeld-Maue
approximation \cite{Fu, ZM, OMW}.

In the calculation of amplitudes of the high-energy processes the
eikonal approximation is often used. The corresponding wave
functions and Green functions differ, generally speaking, from the
quasiclassical ones and have a narrower region of applicability.
Therefore, the use of the eikonal approximation without a special
consideration of its applicability may lead to incorrect results.
As an example, we show this in section 4 for the problem of
small-angle scattering of a charged particle in an external field.
In this section we present also the consequent derivation of the
expression for the scattering amplitude using the quasiclassical wave
function. In particular case of a Coulomb potential the use of the
eikonal wave function instead of the quasiclassical one leads to the
incorrect result.

In section 5 the quasiclassical Green function obtained is used in
the calculation of the amplitude of forward elastic scattering of a
photon in the atomic electric field (forward Delbruck scattering). As
shown in \cite{MS2}, at this calculation it is necessary to take into
account the correction to the quasiclassical Green function. This
correction should be taken into account in the region, where the
eikonal approximation is valid. The contribution of higher
orders of the perturbation theory with respect to the external field
(Coulomb corrections) is determined by the region, where the
eikonal Green function is inapplicable. The real part of the forward
Delbruck scattering amplitude for a screened Coulomb potential is
obtained. It becomes comparable with the amplitude of Compton
scattering already at relatively small energies of a photon. It may
be important for the description of photon propagation in matter.

\section {Green function}
As shown in papers \cite{LM1, LMS1}, in the calculation of
amplitudes of various QED processes it is convenient to use the Green
function of squared Dirac equation $D(\r_2,\r_1|\eps)$, which
is connected to conventional
Green function $G(\r_2,\r_1|\eps)$ by the
relation
\beq\label{g2}
G(\r_2,\r_1 |\,\eps )= \left[
\gamma^{0}(\eps\, -\, V(\r_2))\, -\,\gv\pv_2\, +\, m \right]
D(\r_2,\r_1 |\,\eps )\, ,
\eeq
where $ \gamma^{\mu} $ are the Dirac
matrices, $ \pv = -i\grad $ is the momentum operator, and $V(\r)$ is
the potential.  In the quasiclassical approximation it is possible
to present function $ D (\r_2, \r_1 | \eps) $ as

\beq\label{D}
D(\r_2,\r_1 |\,\eps )= \,\left[1\, -\,\fr{i}{2\eps}(\alv ,
\grad_1 +
\grad_2\,)\right]D^{(0)}(\r_2,\r_1 |\,\eps )\, ,
\eeq
where
\beq\label{D0}
D^{(0)}(\r_2,\r_1 |\,\eps )= \langle\r_2|
\fr{1}{\Q^2\, -\,  H\, +i0}|\r_1\rangle \quad,\quad
 H=\pv^2+2\eps V(\r) -V^2(\r)
\eeq
and $ \Q^2 = \eps^2-m^2 $.
Thus, the problem is reduced to the calculation of the quasiclassical Green
function $ D^{(0)} $ of the Klein-Gordon equation with potential $ V (\r) $
(Schrodinger equation with Hamiltonian $ H $).

Let us pass in function $ D^{(0)} (\r_2, \r_1 | \, \eps) $ from variables $ \r_1 $ and
$ \r_2 $ to variables $ \r_1 $ and $ \r = \r_2-\r_1 $. In terms of these variables the
function $ D_0 $ satisfies to the equation
\beq\label{equ}
[\Q^2-2\Q\phi(\r_1+\r)-\p^2] D^{(0)}(\r+\r_1,\r_1 |\,\eps )
= \delta(\r)\ ,
\eeq
where $ \phi = \lambda V-V^2/2\Q $, $ \lambda = \eps/\Q $, $ \p = -i\grad_r $.
In the ultrarelativistic case $ \lambda = + 1 $ for $ \eps > 0 $ and
$ \lambda = -1 $ for $ \eps < 0 $.

We seek for a solution to this equation in the form
\beq\label{DF}
D^{(0)}(\r+\r_1,\r_1 |\,\eps )=-\fr{\exp(i\Q r)}{4\pi r} F(\r,\r_1)
\ .
\eeq
Note, that the factor in front of $ F $ in \eq{DF} is a Green
function of the equation \eq{equ} for $\phi=0$.  The function $
F $ satisfies the equation \beq\label{Fgen}
\left[i\fr{\partial}{\partial
r}-\phi(\r_1+\r)\right] F= -\fr{1}{2\Q}r\Delta (F/r)
\eeq
with the boundary condition $ F (\r = 0, \r_1) = 1 $.  For further
needs we define an effective impact parameter $ \rho_* $ of a
rectilinear trajectory $ \Gamma $, connecting point $ \r_1 $ and $
\r_2 $, as
\beq
\label{rhoeff}
\rho_*=\min_{\vx\in\Gamma}\fr{|\phi(\vx)|}{|\grad_\perp\phi(\vx)|}\
.
\eeq
Let this minimum is achieved at some point $ \vx = \r_* $. The
necessary condition $ \Q\rho_*\gg 1 $ for the applicability of the
quasiclassical approximation is assumed to be fulfilled.
Introducing the notation $ a_{1,2} = | \r_{1,2} -\r_* | $,
we have two overlapping regions:
\beqn\label{cond}
&&1)\ \mbox{min}(a_1,a_2) \ll \Q\rho_*^2  \\
&&2)\ \mbox{min}(a_1,a_2) \gg \rho_*\
.  \nonumber
\eeqn
In the first region, the r.h.s. of the equation \eq{Fgen} is small.
The solution to this equation with zero r.h.s. has the form
\beq\label{Feik0}
F_0=\exp\left[-i r\int_0^1 \phi(\r_1+x\r)dx\right]\ ,
\eeq
which corresponds to the eikonal approximation. For the calculation of the first
correction to $ F_0 $ we search for a solution of the equation \eq{Fgen} in the
form $ F = F_0 (1 + g) $ and neglect $ g $ in r.h.s.. As a result, we obtain the
following equation for $ g $:
\beq
 2i\Q\fr{\partial}{\partial
r}g=\phi^2(\r_1+\r)+i r\int\limits_0^1 dx\, x^2
\Delta_1\phi(\r_1+x\r)+r^2\left[\int\limits_0^1 dx \, x
\grad_{1\perp}\phi(\r_1+x\r)\right]^2 \ ,
\eeq
where the index $ 1 $ at the derivative designates differentiation
over $ \r_1 $, $ \grad_{1\perp} $ is a component of a gradient,
perpendicular to $ \r $, $ \Delta_1 = \grad_1^2 $.  Integrating over
r we obtain for $ g $
\beqn
g&=&\fr 1{2\Q} \left[r^2\int\limits_0^1
dx\, x(1-x)\Delta_1\phi(\r_1+x\r) -i r\int\limits_0^1 dx\,
\phi^2(\r_1+x\r)-\right. \nonumber\\ &-&\left.2ir^3\int\limits_0^1 dx
(1-x) \grad_{1\perp}\phi(\r_1+x\r) \int\limits_0^x dy \, y
\grad_{1\perp}\phi(\r_1+y\r) \right]
\eeqn
Finally, up to the first correction,
the Green function $ D^{(0)} $ in the first region
reads
\beqn\label{DI}
D^{(0)}(\r_2,\r_1|\eps)&=&-\fr 1{4\pi r}\exp\left[i\Q r-i
\lambda r\int_0^1 V(\r_1+x\r)dx\right]\times\\
&\times&\left\{1+\fr 1{2\Q} \left[\lambda r^2\int\limits_0^1 dx\,
x(1-x)\Delta_1 V(\r_1+x\r) -\right.\right. \nonumber\\
&-&\left.\left.2ir^3\int\limits_0^1 dx
(1-x) \grad_{1\perp}V(\r_1+x\r) \int\limits_0^x dy \, y
\grad_{1\perp}V(\r_1+y\r) \right]\right\}\ .\nonumber
\eeqn
In this expression we put $ \lambda^2 = 1 $ assuming $ \eps\gg
m $. Substituting \eq{ff00} in
\eq{D}, we find the expression for function $ D $ in the first region
\beqn\label{DI1}
&&D(\r_2,\r_1|\eps)=-\fr 1{4\pi r}\exp\left[i\Q r-i
\lambda r\int_0^1 V(\r_1+x\r)dx\right]\times\\
&&\times\left\{1-\fr r{2\Q}\int\limits_0^1 dx\, \alv\grad_{1}
V(\r_1+x\r)
+\fr  {\lambda r^2}{2\Q}\int\limits_0^1 dx\,
x(1-x)\Delta_1 V(\r_1+x\r) -\right. \nonumber\\
&&-\left.\fr{ir^3}{\Q}\int\limits_0^1 dx
(1-x) \grad_{1\perp}V(\r_1+x\r) \int\limits_0^x dy \, y
\grad_{1\perp}V(\r_1+y\r)\right\}\ .\nonumber
\eeqn

In the second region the r.h.s. of the equation
\eq{Fgen} is not small. Using spherical coordinates, we can rewrite
\eq{Fgen} as
\beq\label{F2}
\left[i\fr{\partial}{\partial
r}-\phi(\r_1+\r)-\fr{\lv^2}{2\Q r^2}\right] F=
-\fr{1}{2\Q}\fr{\partial^2}{\partial r^2}F\ ,
\eeq
where $ \lv^2 $ is the angular momentum operator squared. In this
equation $ \r $ is a free variable. We are interested in a value of
function $ F (\r, \r_1) $ at $ \r = \r_2-\r_1 $. It is convenient to
direct the polar axis along $ \r_2-\r_1 $. The term in the l.h.s. of
\eq{F2} containing $ \lv^2 $ should be taken into account for $ r\ge
a_1 $.  It can be checked by applying the operator $ \lv^2 $ to the
eikonal function \eq{Feik0}. When calculating the function
$F(\r_2-\r_1, \r_1) $ it is enough to consider a narrow region of
polar angles of vector $ \r $ where $ \theta\sim \rho_*/a_1\ll 1
$. The r.h.s. of \eq{F2} is small. We seek for a solution to this
equation in the form
\beq\label{af} F=\mbox{e}^{iA} {\cal F}\ , \quad
A=\left(\fr 1r -\fr 1{a_1}\right)\fr {\lv^2}{2\Q} \eeq
Substituting \eq{af} in \eq{F2}, we obtain
$\cal F$ \beq\label{calF}
\left[i\fr{\partial}{\partial r}-\tilde{\phi}\right]{\cal F}=
-\fr{1}{2\Q}\left[\fr{\partial^2}{\partial r^2}
-\fr{i \lv^2}{\Q r^2}\left(\fr{\partial}{\partial r}-\fr 1r
-\fr{i\lv^2}{4\Q r^2}\right) \right] {\cal F} \ ,
\eeq
where
\beq
\tilde{\phi}=\mbox{e}^{-iA} \phi(\r_1+\r) \mbox{e}^{iA}\ .
\eeq
We are going to solve this equation up to the first correction in the
parameter $ 1/\Q\rho_* $.
For this purpose we should keep only two terms in the expansion of
$ \tilde {\phi} $ in terms of commutators of the operator $ A $ and
$ \phi $: $ \tilde {\phi} \approx \phi-i [A, \phi] $. In the first
approximation it is possible to neglect r.h.s. of \eq{calF} and to
replace $ \tilde {\phi} $ by $ \phi $. Then the
function $ {\cal F} $ coincides with the eikonal function $ F_0 $ (see
\eq{Feik0}). To find the first correction, we present $ \cal F
$ as $ {\cal F} = F_0 (1 + g_1) $. We obtain  the
following equation for $ g_1 $
\beq\label{f1}
2i\Q \fr{\partial}{\partial r}g_1=
\phi^2(\r_1+\r)-\left(\fr 1r-\fr 1{a_1}\right)
\left[i\lv^2\phi(\r_1+\r)+2r(\lv\phi(\r_1+\r))\int\limits_0^1 dx
\lv\phi(\r_1+x\r)\right]\ .
\eeq
Integrating it over $ r $, we find
\beqn\label{f1r}
g_1&=&-\fr{i r}{2\Q} \Biggl[
\int\limits_0^1 dx\phi^2(\r_1+x\r)
+ir\int\limits_0^1 dx\, x\left(1-\fr r{a_1} x\right)
\Delta_{1\perp}\phi (\r_1+x\r)+\nonumber
\\
&&+ 2r^2 \int\limits_0^1 dx
\left(1-\fr r{a_1} x\right) (\grad_{1\perp} \phi(\r_1+x\r)) \int\limits_0^x
dy\, y  \grad_{1\perp} \phi(\r_1+y\r)\Biggr]\ .
\eeqn
Here $ \Delta_{1\perp} = \grad_{1\perp}^2 $.
With the help of the expansion in spherical functions it can be shown, that at $
\beta\ll 1 $ for arbitrary function $ g (\r) $ the following relation is true with the
appropriate accuracy:
\beq\label{beta}
\exp[-i\beta^2 \lv^2]\, g(\r)\approx
\int\fr{d\q}{i\pi } \mbox{e}^{i q^2}\,  g(\r+2\beta r \q)\ ,
\eeq
where $ \q $ is a two-dimensional vector, perpendicular to $\r$.
Using \eq{af}, \eq{f1r} and \eq{beta}, we obtain in the second region
the following expression for the Green function $ D^{(0)} $ with
the first correction taken into account:
\beqn\label{DII}
D^{(0)}(\r_2,\r_1 |\,\eps )&=&
\fr{i\mbox{e}^{i\Q r}}{4\pi^2r} \int d\q \exp\left[iq^2-i\lambda
r\int_0^1dx V(\r_x)
\right]\times\\
&&\times\left\{ 1-\fr{i r}{2\Q} \Biggl[
i\lambda r\int\limits_0^1 dx\, x\left(1-\fr r{a_1} x\right)
\Delta_{1\perp}V (\r_x)+\nonumber\right.
\\
&&+ \left. 2r^2 \int\limits_0^1 dx
\left(1-\fr r{a_1} x\right) (\grad_{1\perp} V(\r_x))
\int\limits_0^x dy\, y  \grad_{1\perp} V(\r_y)\Biggr]\right\}\ ,
\nonumber
\eeqn
where $ \r_x = \r_1 + x \r + \q x \sqrt {2r (r-a_1) / (\Q a_1)} $,
$ \r = \r_2-\r_1 $. Let us remind, that the derivatives over $ \r_1 $
in this formula must be calculated at fixed $ \r $.

Note that the term proportional to $ \q $ in $ \r_x $ is essential
only in a narrow region $ | x-a_1/r | \sim \rho_*/r\ll 1 $.  Using
this fact, we can eliminate the quantity $ a_1 $ from the formula
\eq{DII}. In order to do this, we present $ \r_x = \R_x + \delta\r_x
$, where
\beq\label{Rx} \R_x=\r_1+x \r+\sqrt{2x(1-x)r/\Q}\q , \quad
\delta\r_x=\sqrt{\fr{2r}{\Q}} \left(x
\sqrt{r/a_1-1}-\sqrt{x(1-x)}\right) \q\ .
\eeq
expand $ V (\r_x) $ in the exponent with respect to $ \delta\r_x $
up to the first term, and replace $ \r_x $ with $ \R_x $ in the rest
of the expression.  After the integration by parts over $ \q $ we
obtain:
\beqn\label{DIIx} D^{(0)}(\r_2,\r_1 |\,\eps )&=&
\fr{i\mbox{e}^{i\Q r}}{4\pi^2r} \int d\q \exp\left[iq^2-i\lambda
r\int_0^1dx V(\R_x) \right]\times\\ &&\times\left\{ 1+\fr{i r^3}{2\Q}
\int\limits_0^1 dx \int\limits_0^x dy (x-y) (\grad_{1\perp}
V(\R_x))(\grad_{1\perp}V(\R_y))\right\}\ .  \nonumber
\eeqn
As mentioned above, the regions of applicability of two formulas
\eq{DI} and \eq{DIIx} are overlapping. At $ \rho_* \ll\mbox {min}
(a_1, a_2) \ll \Q\rho_*^2 $ it is possible to expand  $ V (\R_x) $ in
$ \q $ up to the second term and integrate over $ \q $. After that
the formula \eq{DIIx} turns into \eq{DI} including the terms of the
order of $1/\Q\rho_*$. Therefore, the result can be presented in the
form which is correct everywhere for $\Q\rho_*\gg 1 $:
\beqn\label{DStr} &&D^{(0)}(\r_2,\r_1 |\,\eps )= \fr{i\mbox{e}^{i\Q
r}}{4\pi^2r} \int d\q \exp\left[iq^2-i\lambda r\int_0^1dx V(\R_x)
\right]\times\\
&&\times\left\{ 1
-\fr\lambda{2\Q}\left[2\int\limits_0^1 dx
V(\R_x)-V(\r_1)-V(\r_2)\right] +\right.\nonumber\\
&&
+\left.
\fr{i r^3}{\Q} \int\limits_0^1 dx
\int\limits_0^x dy \left[\sqrt{x(1-x)y(1-y)}-(1-x)y\right]
(\grad_{1\perp} V(\R_x))(\grad_{1\perp}V(\R_y))\right\}\ .  \nonumber
\eeqn
$ \R_x $ is defined in \eq{Rx}. Indeed, in the first region we can
expand $ V (\R_x) $ in $ \q $ up to the second-order terms and to
integrate over $ \q $, which leads to the formula \eq{DI}.
In the second region one should bear in mind, that the main
contribution to integrals comes from a narrow region of $ x $ and $ y
$ close to $ a_1/r $ with $ \delta x\sim \delta y \sim\rho_*/r $.
Therefore
$$
2\left[\sqrt{x(1-x)y(1-y)}-(1-x)y\right]=
x-y-(x-y)^2-\left(\sqrt{x(1-x)}-\sqrt{y(1-y)}\right)^2\approx
x-y\
$$
up to the second-order terms in $ \rho_*/r $. Besides, in this
region the terms linear in $ V $ can be neglected in \eq{DStr}.
Thus, in the second region \eq{DStr} turns into \eq{DIIx}.

Using \eq{D}, we obtain the final expression for the Green function $
D (\r_2, \r_1 | \eps) $
\beqn\label{DStr1}
&&D(\r_2,\r_1 |\,\eps )=
\fr{i\mbox{e}^{i\Q r}}{4\pi^2r} \int d\q \exp\left[iq^2-i\lambda
r\int_0^1dx V(\R_x)
\right]\times\\
&&\times\left\{ 1 -\fr r{2\Q}\int\limits_0^1 dx\, \alv\grad_{1}
V(\R_x) -\fr\lambda{2\Q}\left[2\int\limits_0^1 dx
V(\R_x)-V(\r_1)-V(\r_2)\right] +\right.\nonumber\\
&&
+\left.
\fr{i r^3}{\Q} \int\limits_0^1 dx
\int\limits_0^x dy \left[\sqrt{x(1-x)y(1-y)}-(1-x)y\right]
(\grad_{1\perp} V(\R_x))(\grad_{1\perp}V(\R_y))\right\}\ .  \nonumber
\eeqn
The advantage of the representations \eq{DStr} and \eq{DStr1} is that
they keep their form in any reference frame.

The integration over the variable $\q$ in formulas \eq{DStr}
and \eq{DStr1} can be interpreted as the account for quantum
fluctuations near the rectilinear trajectory between vectors $ \r_1
$ and $ \r_2 $. The integral over $ \q $ converges at
$ q\ls 1 $. Using this fact, we conclude that the
 quantum fluctuations can be neglected, when
$$
\int_0^1dx \sqrt {x (1-x) r/\Q} | \grad_{1\perp} V (\r_1 + x\r) | \ll
\int_0^1dx | V (\r_1 + x\r) | \,
$$
This condition is actually equivalent to the first condition in
\eq{cond} ensuring applicability of the eikonal approximation.

In the formulas \eq{DStr} and \eq{DStr1} the terms containing a
potential in the pre-exponent factor, give the correction
to the quasiclassical Green function. The expressions obtained are
valid when these correction are small. Our results were obtained for
a localized potential. Nevertheless,  they are valid for any
potential if the correction is small, e.g. for a
superposition of localized potentials.

\subsection {Quasiclassical Green function in a central field}

For a spherically symmetric potential the quasiclassical Green
function without corrections can be obtained also from the results of
 \cite{LM1, LM2}, where it was calculated for the case of
almost collinear vectors $ \r_1 $ and $ \r_2 $. In these papers with
the help of quasiclassical radial wave functions the following
expression for the function $ D^{(0)} (\r_2, \r_1 | \, \eps) $ was
obtained at small angle $ \theta $ between vectors $ \r_2 $ and $ -\r_1 $:
\beq\label{Dsmall}
D^{(0)}(\r_2,\r_1 |\,\eps )=\fr{i\mbox{e}^{i\Q(r_1+r_2)}}{4\pi\Q
 r_1r_2} \int\limits_{0}^{\infty}dl lJ_{0}(l\theta)
\exp\biggl\{i\biggl[
\fr{l^2(r_1+r_2)}{2\Q r_1r_2}+2\lambda\delta(l/\Q)+\lambda(\Phi(r_1)+\Phi(r_2))
\biggr]\biggr\} \, .
\eeq
Here
$$
\Phi(r)=\int\limits_{r}^{\infty}\, V(\zeta)d\zeta\, ,\
\delta(\rho)= -\int\limits_{0}^{\infty}\,
 V\left(\sqrt{\zeta^2+\rho^2}\right)d\zeta
\, ,\
\lambda=\eps /\Q \, .
$$

If the angle $ \theta' = \pi - \theta $ between vectors $ \r_1 $ and
$ \r_2 $ is small, then
\beq\label{Dsmall1}
D^{(0)}(\r_1,\r_2 |\,\eps )= \,-\fr{1}{4\pi |\r_1-\r_2|}\exp\{
i\Q|\r_1-\r_2|\, +
i\lambda\mbox{sign}(r_1-r_2)(\Phi(r_1)-\Phi(r_2))\}
\, .
\eeq
For further transformations it is convenient to rewrite the
expression \eq{Dsmall} in the other form, using the identity
$$
\int dl\, l J_{0}(l\theta)g(l^2)=\,\fr{1}{2\pi} \int
d\q\,\exp({i\q\,\vt})\, g(q^2)\, ,
$$
where $ g(x)$ is an arbitrary
function, and $\q$ is a two-dimensional vector.  Let us substitute
this relation to \eq{Dsmall} and make the change of variables
$$
\q\to \sqrt{\fr{2\Q r_1r_2}{r_1+r_2}} \q-\fr{\Q r_1r_2}{r_1+r_2}\vt\,
.
$$
Defining an impact parameter $ \ro $ by a relation $$
\ro=\fr{\r\times[\r_1\times\r_2]}{r^3}\, ,
$$
where $\r=\r_2-\r_1$,
taking into account, that at $ \theta\ll 1 $ the impact parameter $
\ro\approx \vt r_1r_2 / (r_1 + r_2) $ and $ \rho\ll r $, we can
rewrite \eq{Dsmall} as
\beq\label{Dq} D^{(0)}(\r_2,\r_1
|\,\eps )=\fr{i\mbox{e}^{i\Q r}}{4\pi^2r} \int d\q
\exp\left[iq^2-i\lambda r\int_0^1dx V\left(\r_1+x\r
+\q\sqrt{2 r_1r_2/\Q r}\right) \right] \, .
\eeq
Here $ \q $ is a two-dimensional vector perpendicular to vector
$ \r $. This formula was obtained for small angles $ \theta $.
However, it is correct also for $ \theta\sim 1 $, since in this case
the term, proportional to $ \q $ the in argument of
a potential can be neglected and the Green function \eq{Dq} turns
into the eikonal one.  In particular, for $ \theta '\ll 1 $ it
coincides with \eq{Dsmall1}. The expression \eq{Dq} agrees with the
main (without the correction) term of \eq{DStr} since in the main
approximation it is possible to replace $\sqrt{x(1-x)}$ by
$\sqrt{r_1 r_2/r^2}$.

\section {Wave functions in the quasiclassical approximation}

The obtained expressions for the quasiclassical Green function allow
us to find the quasiclassical wave functions with the first
corrections.  The quasiclassical wave functions were found earlier in
papers \cite{Fu, ZM} for the case of Coulomb field
(Furry-Sommerfeld-Maue wave function) and in \cite{OMW} for arbitrary
decreasing central potential. These wave functions were calculated in
the main approximation. In paper \cite{Akhi} the wave functions and
corrections to them were found for arbitrary potential in the eikonal approximation.

To calculate the wave functions, we use the known (see, for example,
\cite{BZP}) relations
\beqn \label{WF} &&\lim_{r_2\to \infty}
D^{(0)}(\r_2,\r_1|\eps)= -\fr{\exp[i\Q r_2]}{4\pi r_2}
\psi^{(-)*}_{\pv_2}(\r_1)\ ,\\
&&\lim_{r_1\to \infty}
D^{(0)}(\r_2,\r_1|\eps)= -\fr{\exp[i\Q r_1]}{4\pi r_1}
\psi^{(+)}_{\pv_1}(\r_2)\nonumber\ .
\eeqn
Here $ \pv_1 = -\Q\r_1/r_1 $, $ \pv_2 = \Q\r_2/r_2 $,
$ \psi^{(+)}_\pv (\r) $
( $ \psi^{(-)}_\pv (\r) $) denotes a solution of the Klein-Gordon equation containing
at the infinity a plane wave with momentum $ \pv $ and a diverging
(converging) spherical wave. From \eq{DStr} and \eq{WF} we obtain for
the quasiclassical wave function of the Klein-Gordon equation with
the first correction taken into account \beqn\label{psi1} &&
\psi^{(\pm)}_{\pv}(\r)=\pm \int \fr{d\q}{i\pi} \exp\left[i\pv\r\pm
iq^2\mp i\lambda \int_0^\infty dx V(\r_x) \right]\times\\
&&\times\left\{ 1
+\fr\lambda{2\Q}V(\r) \pm
\fr{i }{\Q} \int\limits_0^\infty dx
\int\limits_0^x dy \left[\sqrt{xy}-y\right]
(\grad_{\perp} V(\r_x))(\grad_{\perp}V(\r_y))\right\}\ ,
\nonumber\\
&& \r_x=\r\mp \pv x/\Q+\q\sqrt{2x/\Q}\ ,\quad
\Q=|\pv|\ .
\nonumber
\eeqn
Here $ \q $ is a two-dimensional vector perpendicular to $ \pv $,
$ \grad_{\perp} $ is the component of a gradient, perpendicular to $
\pv $.  Similarly, for the Dirac equation the quasiclassical wave
function with the first correction is obtained from \eq{DStr1}:
\beqn
\label{Dirac} && \Psi^{(\pm)}_{\pv}(\r)= \pm\int \fr{d\q}{i\pi}
\exp\left[i\pv\r\pm iq^2\mp i\lambda \int_0^\infty dx
V(\r_x) \right]\left\{ 1 \mp \fr 1{2\Q}\int\limits_0^\infty dx\,
\alv\grad V(\r_x)+\right.\nonumber
\\
&& +\left.\fr\lambda{2\Q}V(\r) \pm
\fr{i }{\Q} \int\limits_0^\infty dx \int\limits_0^x dy
\left(\sqrt{xy}-y\right) (\grad_{\perp}
V(\r_x))(\grad_{\perp}V(\r_y))\right\} u^\lambda_{\pv}\ ,
\eeqn
where $ u^\lambda_\pv $ is a free positive-energy ($ \lambda = 1 $) and
negative-energy ($ \lambda = -1 $) Dirac spinors with momentum
$\pv$. In the case, when $ V (\r_x) $ can be expanded in $\q$, the formula
\eq{Dirac} turns into the wave function in the eikonal approximation
with correction being in accordance with the result of \cite{Akhi}.

\section {Scattering of charged particles}

In this section we use the obtained quasiclassical Green function for the
calculation of the small-angle scattering amplitude of the high-energy charged
particle in localized potential. In the case of a particle with spin
zero the scattering amplitude has the form:
\beq\label{fgen}
f(\pv_2,\pv_1)=-\fr{\Q}{2\pi}\int d\r
\psi^{(-)*}_{\pv_2}(\r) \phi(\r) \mbox{e}^{i \pv_1 \r}\ .
\eeq
Recollect, that $ \phi = \lambda V -V^2/2\Q $.

For the sake convenience we put the origin of the reference frame at
the point of potential maximum and pass to cylindrical coordinates
with axis $ z $ along the vector $\p_2$. Substituting \eq{psi1} in
\eq{fgen}, we obtain with the account  of the first correction
\beqn\label{f}
f&=&\fr{i\Q}{2\pi^2}\int\limits_{-\infty}^{\infty}
dz\int d\ro \int d\q \exp\left[-iQ_z z-i\vq_\perp \ro + i \q^2
-i\lambda\int_0^\infty dx V(x+z,\ro_x)\right]
\times\\
&&
\times\lambda V(z,\ro) \left[ 1 +\fr{i }{\Q} \int\limits_0^\infty dx
\int\limits_0^x dy \left[\sqrt{xy}-y\right]
(\grad_{\rho} V(x+z,\ro_x))(\grad_{\rho}V(y+z,\ro_y))\right]\
,\nonumber
\eeqn
where $ \ro_x = \ro + \sqrt {2x/\Q} \, \q $, $ \vq = \p_2-\p_1 $.

We demonstrate first that in the scattering problem not the eikonal wave
function but the quasiclassical one should be used. For this
purpose we calculate the amplitude in the main approximation, that
corresponds to the replacement of the factor in square brackets of
\eq{f} by $1$.  We split the region of integration over $ z $ into
two: $ (-\infty, 0) $ and $ (0, \infty) $. In the region from zero to
infinity it is possible to neglect quantum fluctuations, which
means the replacement $ \ro_x \to \ro $ and then to integrate over
$ \q $.  Finally,
 the contribution of this region reads
\beq f_+=
-\fr{\Q}{2\pi}\int\limits_{-\infty}^{\infty} dz\int d\ro
\exp\left[-iQ_z z-i\vq_\perp \ro
-i\lambda\int_0^\infty dx V(x+z,\ro)\right]
\lambda V(z,\ro)\ .
\eeq
Now, integrating by parts over $ z $ with the help of the relation
\beq\label{tozh}
\lambda
V(z,\ro)\exp\left[-i\lambda\int_0^\infty dx
V(x+z,\ro)\right]=-i\fr{\partial}{\partial z}\left\{
\exp\left[-i\lambda\int_0^\infty dx V(x+z,\ro)\right]-1
\right\}  \ ,
\eeq
we obtain
\beqn
f_+&=&-\fr{i\Q}{2\pi} \int d\ro \exp[-i\vq_\perp\ro]
\left(\exp\left[
  -i\lambda\int_{0}^\infty dx V(x,\ro)
\right]-1-\right.\\
&&\left. -i Q_z \int\limits_0^\infty dz \,  \exp[-iQ_z z]
\left(\exp\left[-i\lambda\int_{0}^\infty dx V(x,\ro)
\right]-1\right)
\right) \nonumber
\ .
\eeqn
The main contribution to the integral over $ z $ in this formula
comes from $ z\sim\rho $. The relative magnitude of the contribution
proportional to this integral, in comparison with the first term
 is $ Q_z \rho\sim Q_z/Q_\perp\ll 1 $, therefore in the main
approximation this contribution can be neglected.  In the region from
$ -\infty $ up to zero in the main approximation we can
replace $ x $ with $ | z | $ in $ \ro_x $. Performing the shift $
\ro\to \ro- \q \sqrt {2 | z | /\Q} $ we obtain
\beqn f_-&=&
\fr{i\Q}{2\pi^2}\int\limits_{-\infty}^0 dz\int d\ro \int d\q
\exp\left[-i(Q_z-\vq_\perp^2/2\Q)\, z -i\vq_\perp \ro + i
(\q+\vq_\perp\sqrt{|z|/2\Q})^2- \right.\nonumber \\ &&\left.
-i\lambda\int_0^\infty dx V(x+z,\ro)\right] \lambda V(z,\ro-\q
\sqrt{2|z|/\Q})\ .
\eeqn
Note, that at $ \Q\rho\gg 1 $ at any $ z $ the condition $ \sqrt {2 |
z | /\Q} \ll \max (| z |, \rho) $ holds, which allows us to neglect
the term proportional to $ \q $ in the argument of the potential. In
the  small-angle approximation $ Q_z = \vq_\perp^2/2\Q $, that is,
the term $\propto z $ in the exponent vanishes. Taking the
integral over $ \q $, and then over $ z $ with the use of
the identity \eq{tozh}, we find
\beq f_-=-\fr{i\Q}{2\pi} \int d\ro
\exp[-i\vq_\perp\ro] \left(\exp\left[-i\lambda\int_{-\infty}^\infty
dx V(x,\ro) \right]-\exp\left[-i\lambda\int_{0}^\infty dx V(x,\ro)
\right]\right)\ .
\eeq
Combining $ f_+ $ and $ f_- $, we obtain the expression known as the
scattering amplitude in the eikonal approximation:
\beq\label{mf}
f=-\fr{i\Q}{2\pi} \int d\ro \exp[-i\vq_\perp\ro]
\left(\exp\left[-i\lambda\int_{-\infty}^\infty dx V(x,\ro)
\right]-1\right)\ .
\eeq
Usually this result is derived with the use of the eikonal wave
function in the whole region of integration over $ z $ and neglecting
the term $ Q_z z $ in the exponent. It follows from our consideration
 that at arbitrary momentum transfers both these approximations,
generally speaking, are incorrect. For example, consider a screened
Coulomb potential with the radius of screening $ r_c $. The main
contribution to the amplitude comes from the region $ z\sim r_c $,
$\rho\sim 1/Q_\perp$. If  the quantity $ Q_z z\sim Q_\perp^2 r_c/\Q
$ is not small in this region as compared to unity, the term with
$\q$ in $\ro_x$ can not be neglected, since $ \rho\sim 1/Q_\perp
\ls\sqrt{r_c/\Q}$.  Thus, the eikonal wave function becomes
inapplicable in the region of the main contribution to the amplitude.
Additionally, the term $ Q_z z $ in the exponent can not be
neglected. Keeping this term and still using the eikonal wave
function  leads under condition of $ Q_\perp^2 r_c/\Q\gs 1 $ to wrong
result for the scattering amplitude. In particular, acting so,
one can not reproduce a well known result(Ratherford formula) in the
limit $r_c\to\infty$ (unscreened Coulomb potential). Thus, the
formula \eq{mf} is valid for any $ Q_\perp\ll \Q $, however, its
correct derivation can not be done with the use of the eikonal
wave function.

Let us pass now to the calculation of scattering amplitude with
the first correction.  For the sake convenience we first set the
lower limit of integration over $ z $ to $-L$ , and then take the
limit $L\to\infty$.  Using the identity
\beq \left[
\lambda V(z,\ro)+i\fr{\partial}{\partial z}+
\lambda\int\limits_0^\infty dx \fr{\q}{\sqrt{2\Q x}}\grad_\rho
V(z+x,\rho_x) \right] \exp\left[-i\lambda\int_0^\infty dx
V(x+z,\ro_x)\right]=0 \ ,
\eeq
it is possible to perform the
integration by parts over $z$ in the first term of square brackets of
\eq{f}. As a result, the expression for $f$ acquires the form
$f=f_0+f_1$, where
\beqn\label{ff0} f_0&=&
-\fr{\Q}{2\pi^2}\lim_{L\to\infty} \int\!\! d\ro \int\!\! d\q
\exp\left[i\fr{\vq_\perp^2}{2\Q} L -i\vq_\perp \ro + i \q^2
-i\lambda\int_{-L}^\infty\!\! dx V(x,\ro_{x+L})\right]\ ,
\\
\label{ff1} f_1&=&
\fr{i\Q}{2\pi^2}\int\limits_{-\infty}^{\infty} \!\! dz\int\!\! d\ro
\int\!\! d\q \exp\left[-i\fr{\vq_\perp^2}{2\Q} z-i\vq_\perp \ro + i
\q^2 -i\lambda\int_0^\infty\!\! dx V(x+z,\ro_x)\right]
\times\nonumber\\
&&
\times \left[ \fr{i\lambda }{\Q}V(z,\ro)
\int\limits_0^\infty dx \int\limits_0^x dy \left(\sqrt{xy}-y\right)
(\grad_{\rho} V(x+z,\ro_x))(\grad_{\rho}V(y+z,\ro_y)) \right. +\\
&&
+\left.\fr{\vq_\perp^2}{2\Q}-\lambda \int\limits_0^\infty
dx \fr{\q}{\sqrt{2\Q x}}\grad_\rho V(z+x,\rho_x)\right]\nonumber\ .
\eeqn
The term independent of potential and vanishing at $ \vq_\perp\not =
0 $, is omitted in expression for $f_0$.  We will take it into
account explicitly in the final expression for the amplitude.  In
order to find the limit $L\to\infty$ in function $f_0$, we make the
shifts $ \ro\to \ro - \q\sqrt {2L/\Q} $ and $ \q\to \q- \vq_\perp\sqrt
{L/2\Q} $.  After that the calculation of this limit  and integration
over $\q$ become elementary. With the correction
of the order of $ Q_\perp/\Q $ taken into account we obtain
\beqn\label{ff00} f_0&=&-\fr{i\Q}{2\pi}
\int d\ro \exp\left[ -i\vq_\perp \ro-i\lambda\int_{-\infty}^\infty dx
V(x,\ro) \right] \left[
1+\fr{i\lambda}{2\Q}\int\limits_{-\infty}^\infty dx\, x
\vq_\perp \grad_\rho V(x,\ro)
\right]\ .
\eeqn
The integral over $\ro$ in the expression for $f_1$
\eq{ff1} converges at $ \rho\sim 1/Q_\perp$.  The quantum
fluctuations are important only at $z<0$ and $x$ close to $-z$. Since
the quantity $f_1$ is small being a part of the correction, $ \ro_x $
in the formula \eq{ff1} can be replaced by $\ro_{| z |}$.  Besides
the factor $V(z,\ro)$ can be replaced by $V(z,\rho_{|z|})$ because
the difference between these two is small at any $z$.  The integral
over $z$ from zero to infinity converges at $z\ls\rho$, so
$ z\vq_\perp^2/2\Q \sim \rho\vq_\perp^2/\Q\sim Q_\perp/\Q\ll 1 $.
Therefore, it is possible to replace in the exponent $
z\vq_\perp^2/2\Q $ by $-|z|\vq_\perp^2/2\Q$. After integrating by
parts over $\q$ the term proportional to $\q$  this vector enters
everywhere, except the exponent, only in the argument of potential.
Then the expression for $f_1$ can be transformed to the form
\beq
f_1=\int\limits_{-\infty}^\infty dz\int d\ro\int \fr{d\q}{i\pi}
\exp\left[ i\fr{\vq_\perp^2}{2\Q} |z| -i\vq_\perp \ro +i\q^2 \right]
g(z,\ro+\sqrt{2|z|/\Q}\, \q)\
\eeq
with some function $g$.  If we perform the shift
$\ro\to\ro-\q\sqrt{2|z|/\Q}$ the integral over $\q$ becomes
elementary. As a result we have
\beq \label{ff1a}
f_1=\int\limits_{-\infty}^\infty dz\int d\ro
\exp\left[-i\vq_\perp \ro \right] g(z,\ro)\ .
\eeq
Finally, we obtain
the following expression for $f_1$
\beqn \label{ff1b}
f_1&=&-\fr 1{2\pi}\int d\ro
\int\limits_{-\infty}^\infty dz \exp\left[-i\vq_\perp \ro
-i\lambda\int_0^\infty dx V(x+z,\ro)\right]\times\\
&&
\times
\left[
i\lambda V(z,\ro)
\int\limits_0^\infty dx \int\limits_0^x dy \left(\sqrt{xy}-y\right)
(\grad_{\rho} V(x+z,\ro))(\grad_{\rho}V(y+z,\ro))\right.  +\nonumber\\
&&
+\left.
\fr 12\int\limits_0^\infty dx \int\limits_0^\infty dy
\left(1-\sqrt{x/y}\right)
(\grad_{\rho} V(x+z,\ro))(\grad_{\rho}V(y+z,\ro))
\right]\nonumber
\eeqn
When deriving this formula we have integrated by parts over $\ro$ the
term proportional to $\vq_\perp^2$. As can be checked, the
integrand in \eq{ff1b} is a total derivative over $z$ and the
integration over $z$ becomes trivial.  Combining obtained
expression with \eq{ff00}, and integrating by parts over $\ro$ the
term proportional to $\vq_\perp$, we obtain the scattering amplitude
with the first correction
\beqn\label{fff} f&=&-\fr{i\Q}{2\pi}
\int d\ro \exp\left[ -i\vq_\perp \ro\right] \Biggl\{
\exp\left[-i\lambda\int_{-\infty}^\infty dx V(x,\ro)
\right]-1+
\\
&&+ \exp\left[-i\lambda\int_{-\infty}^\infty dx V(x,\ro)
\right]\left[
\fr{\lambda}{2\Q}\int\limits_{-\infty}^\infty dx\, x
\Delta_\rho V(x,\ro)-\right.\nonumber  \\
&&\left.-\fr{i}{\Q} \int\limits_{-\infty}^\infty dx \int\limits_{-\infty}^x dy
y(\grad_{\rho}V(x,\ro))(\grad_{\rho}V(y,\ro))\right]
 \Biggr\}\ .
\nonumber
\eeqn
Here $ \Delta_\rho = \grad_\rho^2 $.
Using the wave function \eq{Dirac}, it is possible to show (see, for
example, \cite{BK}), that the small-angle scattering amplitude
for particles with spin $1/2$ coincides with amplitude \eq{fff} for
spin-zero particles including the terms of the order of $Q_\perp/\Q$.

The obtained amplitude has correct properties
with respect to shifts. It follows from \eq{f}, that after the
replacement $V(\r)\to V(\r+\r_0)$ the amplitude $f(\p_2,\p_1)$
acquires the factor $\exp[i\vq\r_0]$.  The amplitude
\eq{fff} obviously has this property with respect to the shift in
the direction, perpendicular to $\p_2$, i.e., to $z$-axis. Let us
consider now the shift along the $z$-axis: $ V (z, \ro) \to V
(z + z_0, \ro) $. Performing the change of variables $ x\to x-z_0 $
and $ y\to y-z_0 $ in the integrals over $ x $ and $ y $ , we obtain
the additional term proportional to $ z_0 $:
\beqn
\delta f&=&\fr{iz_0}{4\pi} \int d\ro \exp\left[ -i\vq_\perp
\ro-i\lambda\int_{-\infty}^\infty dx V(x,\ro) \right]\times\\
&&\times \left[ \lambda\int\limits_{-\infty}^\infty dx\, \Delta_\rho
V(x,\ro)-i\left(\int\limits_{-\infty}^\infty dx
\grad_{\rho}V(x,\ro)\right)^2)\right]\nonumber\\
&=&-\fr{z_0}{4\pi} \int d\ro \exp\left[
-i\vq_\perp \ro]\Delta_\rho
\left(\exp[-i\lambda\int_{-\infty}^\infty dx
V(x,\ro) \right]-1\right)\nonumber\ .
\eeqn
Integrating by parts over $ \ro $ and substituting
$ \vq_\perp^2/2\Q\to Q_z $, we find for $ Q_z Z_0\ll 1 $ that $ f +
\delta f\approx f \exp (iQ_z z_0) $. Thus, the correct
transformation properties of the scattering amplitude with respect
to shifts holds with the same accuracy, as formula \eq{fff} itself.

For the potential satisfying the condition $ V (z, \ro) = V
(-z, \ro) $ the expression \eq{fff} agrees with
that obtained in \cite{Akhi} in the eikonal approximation with the
account for the correction. However, as explained above, the
formula \eq{fff} holds even when the eikonal approximation is
inapplicable.

\section {Delbruck scattering}

The process of coherent photon scattering in the electric field of
atoms via virtual electron -positron pairs (Delbruck scattering)
has been intensively investigated both theoretically, and
experimentally (see, e.g., the review \cite{MShu}). In this
section we consider the forward Delbruck scattering in a
screened Coulomb field as one more example of the application of
the quasiclassical Green function obtained. The photon energy
$\omega$ is assumed to be large in comparison with the electron mass
$m$.  According to the optical theorem, the imaginary part of this
amplitude is proportional to the cross section of electron-positron
pair production by a photon in the field of atom.  To describe the
propagation of light in matter it is also necessary to know the real
part of this amplitude.

For the calculation of amplitudes of Delbruck scattering with the
help of Green functions one is forced to use different
approximations for these functions for different momentum transfers
$\vd=\k_2-\k_1$ ($\k_{1},\k_2$ are the momenta of initial and
final photons). At $ \Delta \sim \omega $ the quasiclassical
approximation is inapplicable, since the basic contribution to the
amplitude is given by the orbital moment $ l\sim\omega/\Delta\sim 1
$.  For such momentum transfers the amplitude was calculated in
\cite{MSh}. In region $ \omega\gg\Delta \gg m^2/\omega $ it is
possible to use the quasiclassical Green function in the main
approximation \cite{MS1, LMS4}.  The calculation of Delbruck
scattering amplitude in Coulomb field at $ \Delta\ls
 m^2/\omega $ required a special consideration (see \cite{CW2, MS2}).
In this case the contribution to the amplitude is given by the impact
parameters $\rho$ up to $ \rho\sim\omega/m^2 $.  At such impact
parameters it is necessary to take into account a correction to the
quasiclassical Green function \cite{MS2}.  For the screened Coulomb
potential with $ \omega/m^2\gg r_c $ ( $ r_c $ is the radius of
screening, in Thomas-Fermi model $ r_c\sim(M\alpha)^{-1} Z^{-1/3}\gg 1/m$)
the contribution to the amplitude is given
by the impact parameters $ \rho\ls r_c\ll\omega/m^2 $. In this case,
the quasiclassical Green function without corrections  can be used
for arbitrary $ \Delta\ll \omega $ \cite{LM1, LM2}. However, if $
\Delta, \, r_c^{-1} \ls m^2/\omega $, the correction is
very important.  Moreover, the expression for the forward scattering
amplitude ($\Delta=0$), obtained with the use of the quasiclassical
Green function without corrections is, strictly speaking, indefinite.
Below we derive the amplitude of forward Delbruck scattering at
arbitrary ratio between $ r_c $ and $ \omega/m^2 $.

As shown in \cite{LM1}, the amplitude of forward Delbruck
scattering for high photon energy can be presented as follows
\beqn\label{amplitude}
M&=&i\alpha\int\limits_0^\omega d\eps
\int\limits d\r_1 d\r_2\exp [i\k(\r_1-\r_2)]
\times\\
&&
\times\mbox{Sp}\biggl[(2\e^{*}\pv_2-\hat e^{*}\hat k)
D(\r_2 ,\r_1 |\omega -\eps)\biggr]
\biggl[ (2\e\pv_1+\hat e\hat k) D(\r_1 ,\r_2 |-\eps)\biggr]\nonumber
\eeqn
where $e,\,k$ are polarization and 4-momentum vectors of a photon,
$ \pv_{1,2} = -i\grad_{1,2} $. In this formula the subtraction from
integrand  of its value at zero external field is assumed to be
done.  Since in the case of central field the amplitude of forward
scattering does not depend on polarization of photon, it is
convenient  to make the substitution
$e_i^*e_j\to(\delta_{ij}-k_ik_j/\omega^2)/2$ in
\eq{amplitude}.

Let us pass in \eq{amplitude} from variables $\r_{1,2}$ to
$$
\r=\r_2-\r_1,\quad \ro=\fr{\r\times[\r_1\times\r_2]}{r^3},\quad
z=-\fr{(\r\r_1)}{r^2}\, .
$$

Since $ \ro\r = 0 $, the integration over $\ro$ is carried
out in a plane, perpendicular to $\r$. The main
contribution to the amplitude comes from the integration region,
where $ r\sim\omega/m^2$, $|z|\sim 1$ and angles between vector $ \r
$ and $ \k $ are of the order $ \theta_r \sim m/\omega \ll 1 $. Due
to the smallness of angles $\theta_r$ it is possible to consider
vector $\ro$ to be perpendicular to $\k$ too. Besides, it is obvious
that the main contribution is given by $ \rho\ls
\mbox{min}(1, \, m^2r_c/\omega)$.

Let us split the region of integration over $ \rho $ into two:
from $0$ up to $\rho_0$ and from $\rho_0$ to $\infty$, where
$m/\omega\ll\rho_0\ll \mbox{min}(1, \, m^2r_c/\omega) $.  In the
first region (at $ \rho < \rho_0 $) the following form of the
quasiclassical Green function can be used
\beqn\label{Dq1}
D(\r_2,\r_1|\,\eps)&=&\fr{i\mbox{e}^{i\Q
r}}{4\pi^2r} \int d\q \left[1+\fr{\alv\q}{\eps}\sqrt{\fr{\Q r}{2 r_1
r_2}}\right] \times \\ &&\times\exp\left[iq^2-i\lambda r\int_0^1dx
V\left(\r_1+x\r +\q\sqrt{2 r_1r_2/\Q r}\right) \right]\, , \nonumber
\eeqn
which we obtained by substituting \eq{Dq} in \eq{D} and performing
some transformations of the term containing $ \alv $-matrix. Namely,
the longitudinal components of gradient, which in comparison with
transverse ones have additional smallness $\rho$, were omitted and
the integration by parts over $\q$ was performed.  Besides, by virtue
of the definition of $\rho_0$, in this region we can neglect
screening and replace $ V (r) $ by a Coulomb potential $ V_c (r)
= -Z\alpha/r $.  After this the integral over $ x $ in the exponent
can be easily taken.

The screening is essential only in the second region, where we can
use the representation \eq{DI1}, that is, the eikonal Green function
with the first correction. The substitution of \eq{DI1} into
\eq{amplitude} results in the cancellation of the
potential-dependent terms in the exponent. Therefore, the
contribution from the second region does not contain any powers of
the potential except the second one, which corresponds to the first
Born approximation.  Besides, due to the phase cancellation we
must take into account the correction to the Green function.  In this
region the contributions to the amplitude rising from the correction
and from the main term in Green function turn out to be of the
same order of magnitude.

Going over to the calculation of the contribution $M_1$ from the
first region, we substitute the Green function \eq{Dq1} for a pure Coulomb
potential into \eq{amplitude} , differentiate and take the trace.
Using the smallness of angles between the vectors $ \r $ and $ \k $
and expanding $ \sqrt {\eps^2-m^2} \approx | \eps | -m^2/2 | \eps |
$, we obtain
\beqn\label{a1}
M_1&=&-\fr{i\alpha}{(2\pi)^4}\int\limits_0^\omega\!\! d\eps
\eps\Q\int\!\! dr\, r^5 d\vt_r
\int\limits_{\rho<\rho_0}\!\!\!d\ro\int\limits_0^1 \fr{dz}{(z(1-z))^3}
\int\!\! d\q_1 d\q_2
\times\\
&&\times
\left[\mbox{Re}\left(\fr{|\ro-\q_1|}{|\ro-\q_2|}\right)^{2iZ\alpha}\!\!\!-1\right]
\exp\left\{i\fr{r}{2}\left[\omega\vt_r^2
-\fr{m^2\omega}{\eps\Q}+\fr{\eps q_1^2+\Q q_2^2}{z(1-z)}
\right]\right\}\times \nonumber\\
&&\times\left[
2\eps\Q[\q_1\q_2-z(1-z)\vt_r^2]
+\fr{\omega}{4(z(1-z))}(\eps\q_1-\Q\q_2,\,\q_1-\q_2)-i\fr{\omega}{r}
\right]\, ,
\nonumber \eeqn
where $ \Q = \omega-\eps $ as well as vectors $ \q_{1,2} $, and $
\vt_r $ are  two-dimensional vectors, perpendicular to $
\k $. Note that the integral over $ z $ in this formula is taken
in the limits from zero to unity. The reason is that outside this
interval the Green function have the eikonal form and the phases
depending on a potential are cancelled. For $ \rho < \rho_0 \ll 1 $
it results in a negligible contribution from the region outside the
interval $ 0\leq z\leq 1 $ compared to the contribution from
this interval.

Let us integrate now over $ \vt_r $, pass from the variables $
\q_{1,2} $ to $ \vq = (\q_1 + \q_2) /2 $ and $ \q = (\q_1-\q_2) /2 $
and make the shift $ \ro\to \ro + \vq $. After that the
integral over $\ro$ acquires the form
\beq\label{intrho}
J=\int\limits_{|
\mbox{\boldmath$\rho$\unboldmath}+\vq|<\rho_0}d\ro
\left[\mbox{Re}\left(\fr{|\ro-\q|}{|\ro+\q|}\right)^{2iZ\alpha}-1\right]\, .
\eeq
The main contribution to the amplitude \eq{a1} comes from the region
$ Q, \, q\sim m/\omega\ll\rho_0 $, where we can neglect $ \vq $
in the limit of integration in \eq{intrho}. To take this integral
\eq{intrho} we subtract and add to the integrand the function
$-2(\al)^2 \, [2\ro\q / (\rho^2 + q^2)]^2 $, which is easily
integrated:
\beq J_1=\int\limits_{\rho<\rho_0}d\ro
\left[-2(\al)^2\left(\fr{2\ro\q}{\rho^2+q^2}\right)^2\right]=
-4\pi(\al)^2 q^2 \left(\ln\fr{\rho_0^2}{q^2}-1\right)\, .
\eeq
In turn, dealing with the difference we can extend the integral over
$\rho$ to infinity owing to the fast convergence of the integral.  To
calculate this integral, it is convenient to multiply the integrand
by
\beq\label{delta}
1\equiv\int_{-1}^{1}dy\,\delta
\left(y-\fr{2\ro\q}{\rho^2+q^2} \right)
=(\rho^2+q^2)\int_{-1}^{1}\fr{dy}{|y|}
\delta((\ro-\q /y)^2 -q^2(1/y^2-1)) \, ,
\eeq
and change the order of
integration over $ \ro $ and $ y $. Integrating over
$ \ro $, we obtain
\beq
J_2=4\pi q^2\int\limits_0^1 \fr{dy}{y^3} \left[\mbox{Re}
\left(\fr{1-y}{1+y}\right)^{iZ\alpha}-1+2(\al)^2 y^2\right]
\eeq
Using the replacement $y=\tanh\tau$ we find for $ J = J_1 + J_2 $
\beq\label{Jf}
J=8\pi q^2
(\al)^2\left[\ln\fr{2q}{\rho_0}-1+\mbox{Re}\psi(1+i\al)+C\right]\,
\eeq
where $ C = 0.577... $ is the Euler constant, $\psi(x)=d\ln \Gamma (x)/dx$.
It is convenient to take the remaining integrals in the following
order:  over $ \vq $, $ \q $, $ r $, $ z $, and $ \eps $.  Finally,
the contribution from the first region reads
\beq \label{a1f}
M_1=i\fr{28\alpha(\al)^2\omega}{9m^2}
\left[\ln\fr{\omega\rho_0}{m}
-i\fr{\pi}{2}-\mbox{Re}\psi(1+i\al)-C-\fr{47}{42}\right]\, .
\eeq
The contribution of the higher orders of the perturbation theory in
the external field (Coulomb corrections)
is given by the term $ -\mbox {Re} \psi (1 + i\al) - C $ in
\eq{a1f} and coincides with the known result \cite{CW2}.  Thus, the
Coulomb corrections are completely determined by the first region,
when the quasiclassical Green function is not reduced to the
eikonal one.

Let us pass to the calculation of the contribution $ M_2 $ from the
second region. Taking the derivatives over $ \r_{1,2} $,
calculating the trace over gamma-matrices and integrating over $
\vt_r $ we come to the following representation for $M_2$:
\beqn\label{a2}
M_2&=&\fr{\alpha}{2\pi\omega}
\int\limits_0^\omega\!\!  d\eps \int\!\! dr\, r^2
\exp\left[-i\fr{\omega r m^2}{2\eps\Q}\right]
\int\limits_{\rho>\rho_0}\!\!\!d\ro\int\limits_{-\infty}^\infty
dz \int\limits_0^1\!\!\!\int\limits_0^1 dx\, dy\times\\
&&\times\left[2y(1-x)
[2\vartheta(x-y)+1]-\fr{\omega^2}{2\eps\Q}\right] [\grad_\rho
V(R_{z-x})]\cdot [\grad_\rho
V(R_{z-y})] \, ,\nonumber
\eeqn
where $R_s = r \sqrt {s^2 + \rho^2} $.  In this formula the terms,
antisymmetric with respect to the substitution $ \eps\to\omega-\eps,
\, z\to 1-z $ , are omitted, since their contribution to the integral
vanishes.  We emphasize, that in contrast to the
first region, in the second one the integration over $ z $ is
performed in infinite limits.  Changing the variables $ z\to z + x $,
$ y\to y + x $ in triple integral over $ x, y, z $ and performing the
integration over $ x $, we have
\beqn\label{a21}
M_2&=&\fr{\alpha}{2\pi\omega} \int\limits_0^\omega\!\!  d\eps
\int\!\! dr\, r^2 \exp\left[-i\fr{\omega r m^2}{2\eps\Q}\right]
\int\limits_{\rho>\rho_0}\!\!\!d\ro\int\limits_{-\infty}^\infty
dz \int\limits_0^1 dy (1-y)\times\\
&&\times
\left[\fr{4}{3}(1-y)^2+2y-\fr{\omega^2}{\eps\Q}\right]
[\grad_\rho V(R_z)]\cdot [\grad_\rho V(R_{z-y})] \, .  \nonumber
\eeqn
At $ r\ll r_c$ the potential $ V (r)\approx -\al/r $. Therefore, it is
convenient to present the contribution \eq{a21} as a sum $ M_2 =
M_2^{c} + \delta M $, where $ M_2^c $ is the value of $ M_2 $ at $ V
(r) = V_c (r) = -\al/r $.

At $ V = V_c $ the integrals over $ r $ and $ \eps $ can be easily
taken, and we find
\beq
M_2^c=-\fr{2i\alpha(\al)^2\omega}{m^2}
\int\limits_{\rho_0}^\infty\!\!\!d\rho \rho^3
\int\limits_{-\infty}^\infty dz \int\limits_0^1 dy (1-y)
\fr{2(1-y)^2/9+y/3-1}{[z^2+\rho^2]^{3/2}[(z-y)^2+\rho^2]^{3/2}}
\, .
\eeq
Using the Feynman parametrization of denominators
$$
\fr{1}{(AB)^{3/2}}=\fr{8}{\pi}\int\limits_0^1 dv\fr{\sqrt{v(1-v)}}{[A
v+B(1-v)]^3}\, ,
$$
we take the integrals over $ \rho $ and $ z $, and then over $ y $
and $ v $ bearing in mind, that $ \rho_0\ll 1 $. Finally,
for this (Coulomb) contribution we obtain
\beq\label{a2cf}
M_2^c=-i\fr{28\alpha(\al)^2\omega}{9m^2}\left(\ln
\fr{\rho_0}{2}+\fr{31}{21}\right)
\eeq
The sum of the contributions \eq{a1f} and
\eq{a2cf} gives the known result for a pure Coulomb potential
\cite{CW2}:
\beq\label{coul}
M_c=i\fr{28\alpha(\al)^2\omega}{9m^2}
\left[\ln\fr{2\omega}{m}
-i\fr{\pi}{2}-\mbox{Re}\psi(1+i\al)-C-\fr{109}{42}\right]\, .
\eeq

In the term $ \delta M $, connected to screening, the lower limit of
integration over $\rho $ can be replaced by zero owing to the
convergence of the integral at $\rho\to 0$.  Using the momentum
representation for potentials
$$
V(r)=\int \fr{d
\pv}{(2\pi)^3}\mbox{e}^{i\pv \r} \tilde{V}(p)\, ,
$$
we take the integrals over $ \ro $ and $ z $. The result of this
integration is proportional to $ \delta (\pv_1-\pv_2) $.  Integrating
over $ \pv_2 $ and over the angles of vector $ \pv_1 $, we find
\beqn
\delta M&=&\fr{\alpha}{2\pi^3\omega} \int\limits_0^\omega\!\!  d\eps
\int\!\! dr\,r \exp\left[-i\fr{\omega r m^2}{2\eps\Q}\right]
\int\limits_0^1 dy (1-y)
\left[\fr{4}{3}(1-y)^2+2y-\fr{\omega^2}{\eps\Q}\right]\times\\
&&\times
\int\limits_0^\infty dp\,
\left[p^4\tilde{V}^2(p)-(4\pi\al)^2\right]\left(\fr{\sin\zeta}{\zeta^3}-
\fr{\cos \zeta}{\zeta^2}\right)  \, ,  \nonumber
\eeqn
where $ \zeta = r p y $. Passing from the variable $ r $ to $
\zeta $, integrating by parts over $p$ and taking the integral over
$ \zeta $, we obtain
\beqn
\delta M&=&\fr{\alpha\omega}{2\pi^3m^2}
\int\limits_0^1\!\!  dx x(1-x)
\int\limits_0^1 dy (1/y-1)
\left[\fr{4}{3}(1-y)^2+2y-\fr{1}{x(1-x)}\right]
\times\\
&&\times
\int\limits_0^\infty dp\,(\partial_pp^4\tilde{V}^2(p))
\left[\fr{1}{\eta}+\fr{1}{2}\left(1-\fr{1}{\eta^2}\right)
\ln\left(\fr{1+\eta}{1-\eta-i0}\right)
\right]  \, ,  \nonumber
\eeqn
where the substitution $ \eps\to \omega x$ is made,
 $ \eta = 2\omega x (1-x) p y/m^2 $. Now we make the change
of variables $ y\to y / (2x (1-x)) $, change the
order of integration over $ x $ and $ y $, and
take the integral over $ x $. Finally, the contribution to the
amplitude of forward Delbruck scattering due to screening reads
\beqn \label{dMfinal}
\delta M&=&-\fr{\alpha\omega}{18\pi^3m^2}
\int\limits_0^\infty dp\,(\partial_pp^4\tilde{V}^2(p))
\int\limits_0^{1/2}\fr{dy}{y}
\left[\fr{1}{\eta}+\fr{1}{2}\left(1-\fr{1}{\eta^2}\right)
\ln\left(\fr{1+\eta}{1-\eta-i0}\right)
\right]\times\nonumber\\
&&\times
\left[(6y^2+7y+7)\sqrt{1-2y}+3y(2y^2-3y-3)
\ln\left(\fr{1+\sqrt{1-2y}}{1-\sqrt{1-2y}}\right)\right]
\, ,
\eeqn
where $ \eta = \omega p y/m^2 $. This formula holds for arbitrary
form of a screened Coulomb potential. In some special cases it can
be essentially simplified.  For the potential
$V(r)=-\al\exp(-\beta r)/r$, when
$\tilde{V}(p)=-4\pi\al/(p^2+\beta^2)$, all integrals in \eq{dMfinal}
can be taken:
\beqn
\dst
\delta M=\fr{4i\alpha(\al)^2\omega}{9m^2}
\left[33-13\tau^2+\fr{3}{2}\tau^4+\fr12\tau(24-13\tau^2+3\tau^4) L+
\fr38(8-9\tau^2+\tau^6) L^2\right]\,;\nonumber\\
\dst
\tau=\sqrt{1+\fr{2im^2}{\omega\beta}}\quad,\quad
L=\ln\left(\fr{\tau-1}{\tau+1}\right)\, .
\eeqn
For more realistic Moliere potential \cite{Mol}, we have
\beqn\label{Moliere}
\dst
\tilde{V}(p)=-4\pi\al\sum_{n=1}^{3}\,\fr{\alpha_{n}}{p^2+\beta_{n}^2}
\, ,\\
\dst
\alpha_{1}=0.1, \quad \alpha_{2}=0.55 ,\quad
\alpha_{3}=0.35 , \quad \beta_{n}=\beta_0 b_n ,\nonumber
\\
\dst \nonumber
b_{1}=6 , \quad b_{2}=1.2 , \quad b_{3}=0.3 ,
\quad \beta_0= mZ^{1/3}/121.
\eeqn
In this case
\beqn \label{dMMol}
\dst
\delta M=-\fr{4i\alpha(\al)^2\omega}{9m^2}
\int\limits_0^{1/2}\fr{dy}{y}
\biggl[(6y^2+7y+7)\sqrt{1-2y}+3y(2y^2-3y-3)
\times\\
\dst
\times
\ln\left(\fr{1+\sqrt{1-2y}}{1-\sqrt{1-2y}}\right)\biggr]
 \left\{\sum\limits_{n\neq
k}\alpha_n\alpha_k\left[1-\fr{2i}{\gamma_n+\gamma_k}-
\fr{2}{\gamma_n^2-\gamma_k^2}\ln\left(\fr{\gamma_n+i}{\gamma_k+i}\right)
\right]+\sum\limits_n\alpha_n^2\fr{\gamma_n}{\gamma_n+i} \right\}
 \,
, \nonumber
\eeqn
where $ \gamma_n = \omega\beta_n y/m^2 $.

As known, the imaginary part of the amplitude of forward Delbruck
scattering  is connected with the total cross section $\sigma$ of the
electron-positron pair production by photon in an external field via
the relation $ \sigma = \mbox{Im} M/\omega $. For the Moliere
potential, using \eq{coul} and \eq{dMMol}, we obtain the following
expression for $\sigma$:
\beqn \dst
\sigma=\fr{28\alpha(\al)^2}{9m^2}\Biggl\{\ln\fr{2\omega}{m}
-\mbox{Re}\psi(1+i\al)-C-\fr{109}{42}-\nonumber\\
\dst
-\int\limits_0^{1/2}\fr{dy}{y}
\left[(\fr67y^2+y+1)\sqrt{1-2y}+\fr37y(2y^2-3y-3)
\ln\left(\fr{1+\sqrt{1-2y}}{1-\sqrt{1-2y}}\right)\right]
\times\nonumber\\
\dst
\times
\left[\sum\limits_{n\neq
k}\alpha_n\alpha_k\left[1-
\fr{1}{\gamma_n^2-\gamma_k^2}\ln\left(\fr{1+\gamma_n^2}{1+\gamma_k^2}\right)
\right]+\sum\limits_n\alpha_n^2\fr{\gamma_n^2}{1+\gamma_n^2}
\right]\Biggr\}\, ,
\eeqn
which agrees with the results, known in literature (see,
e.g. \cite{DBM}) .

Let us discuss now the dependence of the real part of Delbruck
amplitude from the photon energy $\omega$.  When
$\omega/m^2\ll r_c$ the screening can be neglected and
$\mbox{Re}M=\mbox{Re}M_c=14\pi\alpha(\al)^2\omega/9m^2$.
The linear growth of the real part with increasing $\omega$
gradually becomes weaker and for $ \omega/m^2\gg r_c $  we obtain
from \eq{coul} and \eq{dMfinal}
\beq \mbox{Re}
M\approx\fr{\alpha}{2\pi^3} \ln^2\left(\fr{\omega}{m^2r_c}\right)\,
\int\limits_0^\infty
\fr{dp}p\,(\partial_pp^4\tilde{V}^2(p))\, .
\eeq
Here only the term containing a higher
degree of the large logarithm is kept. For the illustration  the real
part of the forward Delbruck amplitude is shown in Fig. 1 for $ Z =
 82 $ and $V(r)=-\al\exp(-m\alpha Z^{1/3} r)/r$ as a function of $
\omega $.

One of the basic mechanisms of elastic scattering of a photon is the
Compton scattering on atomic electrons. For the forward scattering
the amplitude of this process is real and does not depend on $
\omega $: $ M_{Comp} = - 4\pi\al/m $. It is seen in Fig. 1, that
the interference between the amplitudes of Compton and
Delbruck forward scattering should be taken into account already at
relatively small energies.

From the calculation of the Delbruck scattering amplitude we
learn ones more, that the use of the eikonal approximation for the
description of high-energy small-angle scattering processes without
proper ground can lead to incorrect results.  For instance, the
Coulomb corrections in the imaginary part of the forward Delbruck
scattering amplitude (and, therefore, in the total cross section of
pair production) would be completely lost if we used the eikonal
Green function in the calculation.

\newpage

\begin {thebibliography} {99}

\bi {MS1} A.I.Milstein and V.M.Strakhovenko, Phys. Lett {\bf 95 A}, 135 (1983).

\bi{MS2} A.I.Milstein and V.M.Strakhovenko, Zh. \'Eksp. Teor. Fiz.
{\bf 85},14 (1983)  [JETP {\bf 58}, 8 (1983)].

\bi {MS0} A.I.Milstein and V.M.Strakhovenko, Phys. Lett {\bf 90 A},
447 (1982).

\bi{LMS1} R.N.Lee, A.I.Milstein, and V.M.Strakhovenko, Zh. \'Eksp.
Teor. Fiz. {\bf 112},1921 (1997)  [JETP {\bf 85}, 1049 (1997)].

\bi {LMS2} R.N. Lee, A.I. Milstein, V.M.Strakhovenko,
Phys. Rev. {\bf A 57}, 2325 (1998).

\bi {LM1} R.N. Lee, A.I. Milstein, Phys. Lett. {\bf A 198}, 217 (1995).

\bi{LM2} R.N.Lee and A.I.Milstein, Zh. \'Eksp. Teor. Fiz. {\bf 107},
1393 (1995)  [JETP {\bf 80}, 777 (1995)].

\bi {Fu} W. Furry, Phys. Rev. {\bf 46}, 391 (1934)

\bi {ZM} A. Sommerfeld, A. Maue, Ann. Phys. {\bf 22}, 629 (1935)

\bi {OMW} H. Olsen, L.C. Maximon, and H. Wergeland, Phys. Rev. {\bf 106}, 27
(1957)

\bi {Akhi} A.I. Akhiezer, V.F. Boldyshev, N.F. Shul'ga, Teor.  Mat.
Fiz. {\bf 23}, 11 (1975).

\bi {BZP} A.I. Baz', Y.B. Zel'dovich, A.M. Perelomov,
" Scattering, reactions and decays in a non-relativistic
quantum mechanics ", Nauka, Moscow 1971.

\bi {BK} V.N. Baier, V.M. Katkov, Dokl. Akad. Nauk SSSR {\bf 227},
325 (1976).

\bi {MShu} A.I.Milstein and M.Schumacher, Phys. Rep. {\bf 243}, 183 (1994).

\bi {MSh} A.I.Milstein and R.Zh. Shaisultanov, J. Phys. {\bf A 21}, 2941 (1988).

\bi{LMS4} R.N.Lee, A.I.Milstein, and V.M.Strakhovenko, Zh. \'Eksp.
Teor. Fiz. {\bf 116},1 (1999).

\bi {CW2} M.Cheng and T.T.Wu, Phys. Rev. {\bf D 2}, 2444 (1970).

\bi {Mol} G.Z.Moli\`ere, Z. Naturforsch. {\bf 2a}, 133 (1947).

\bi {DBM} H. Davies, H.A. Bethe, and L.C. Maximon, Phys. Rev. {\bf 93}, 788
(1954)

\end {thebibliography}

\newpage

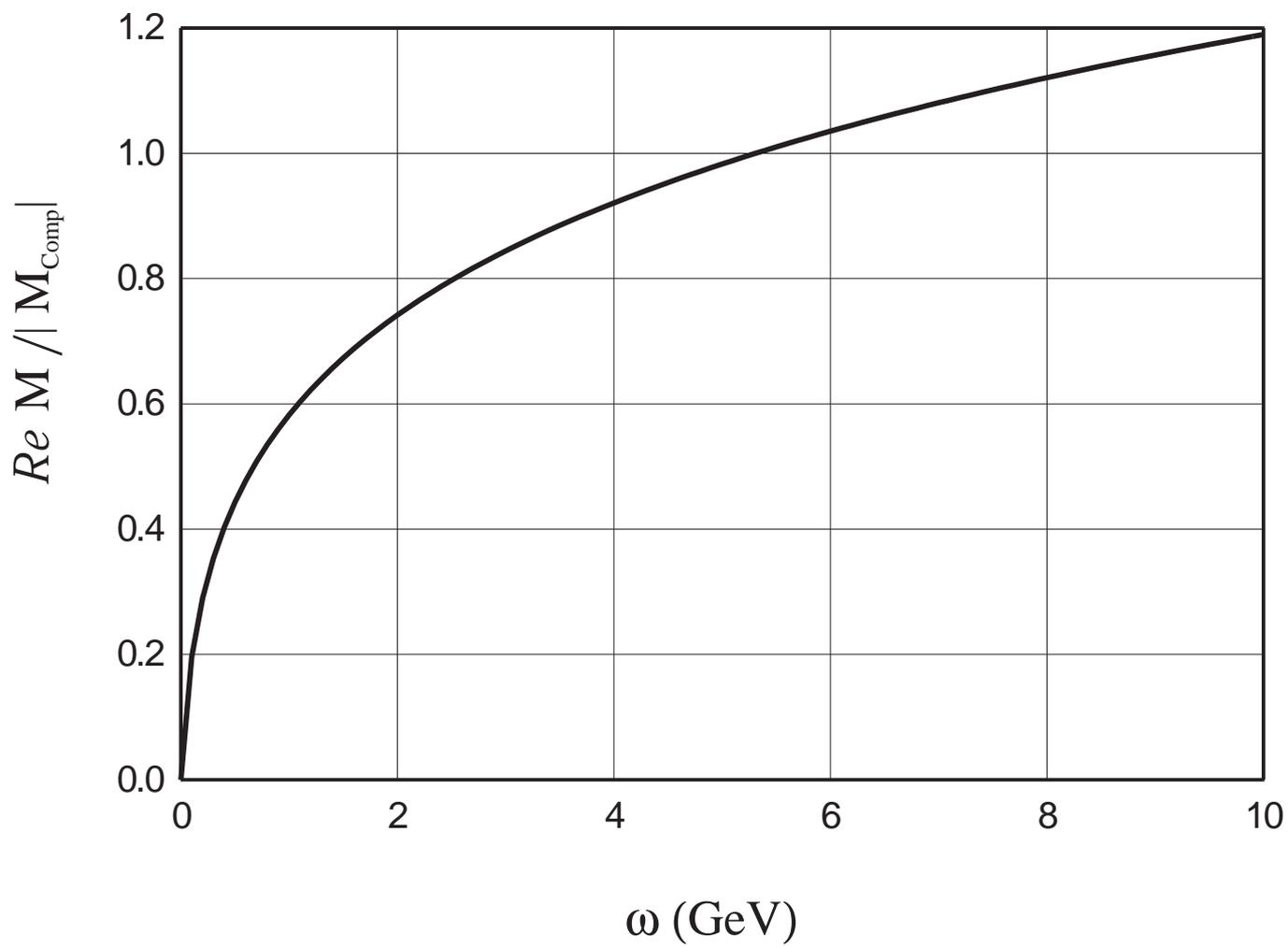
Fig. 1. The real part of the forward Delbruck scattering amplitude
for the potential $ V (r) = -\al\exp (-m\alpha Z^{1/3} r) /r $ in
units of $ 4\pi\al/m $, $ Z = 82 $ as a function of $ \omega $.

\end {document}